\documentstyle[12pt]{article}

\newcommand{\vs}[1]{\vspace{#1 mm}}
\newcommand{\hs}[1]{\hspace{#1 mm}}
\renewcommand{\thefootnote}{\fnsymbol{footnote}}
\newcommand{\EQ}{\begin{equation}}
\newcommand{\EN}{\end{equation}}
\newcommand{\bea}{\begin{eqnarray}}
\newcommand{\ena}{\end{eqnarray}}
\newcommand{\nn}{\nonumber\\}

\begin{document}
\topmargin 0pt
\oddsidemargin 5mm

\begin{titlepage}
\begin{flushright}
OU-HET 280 \\
hep-th/9710218
\end{flushright}

\vspace{10mm}
\begin{center}
{\Large Creation of Fundamental String in M(atrix) Theory}
\vspace{15mm}

{\large 
Nobuyoshi Ohta,\footnote{ohta@phys.wani.osaka-u.ac.jp}
Takashi Shimizu\footnote{simtak@phys.wani.osaka-u.ac.jp} and 
Jian-Ge Zhou\footnote{jgzhou@phys.wani.osaka-u.ac.jp, JSPS postdoctral fellow}
}
\vspace{10mm}
 
{\em Department of Physics, Osaka University, \\
Toyonaka, Osaka 560, Japan}
\end{center}
\vspace{15mm}

\centerline{{\bf{Abstract}}} \vs{5}
The potential between two D4-branes at angles with partially unbroken
supersymmetry is computed, and is used to discuss the creation of a
fundamental string when two such D4-branes cross each other in M(atrix) theory.
The effective Lagrangian is shown to be anomalous at 1-loop approximation,
but it can be modified by bare Chern-Simons terms to preserve
the invariance under the large gauge transformation.
The resulting effective potential agrees with that obtained from the string
calculations. The result shows that a fundamental string is created
in order to cancel the repulsive force between two D4-branes at proper angles.
\end{titlepage}

\newpage
\renewcommand{\thefootnote}{\arabic{footnote}}
\setcounter{footnote}{0} 

Recently the creation of a fundamental string when a D0-brane crosses a
D8-brane in type IIA theory, and the creation of a longitudinal membrane
in M-theory when two longitudinal M5-branes cross each other, have been
studied from various points of view~[1-7].

In the string theory context, by requiring RR charge conservation,
it was first found in~\cite{HW} that a third brane is created when two
certain branes cross. In~\cite{BDG}, the anomaly equation was exploited
to show that when the two branes in question cross each other an energy
level crosses zero and a single particle or hole is created, and this was
interpreted as the creation of an open string or brane.
It was also suggested that the induced charge on the D8-brane worldvolume
indicates the creation of a string when the D0-brane crosses the
D8-brane~\cite{DFK}. The one-loop open string calculation in ref.~\cite{BGL}
revealed that the potential of the D0-D8 system vanishes due to the
cancellation of the forces coming from the dilaton-graviton exchange and
a fundamental string created when two such branes cross.

On the other hand, more and more evidence has accumulated to show that
M-theory could be described in terms of a matrix model -- M(atrix)
theory~\cite{BFSS}. Many consistency checks have been done, including
the calculation of potentials between various D-branes~[9-14]. It is then
natural to study the brane creation in the context of M(atrix) theory,
but only a few papers discuss this phenomenon. The authors of
ref.~\cite{HWu} argued that the effective potential between two M5-branes
is dominated by the contribution from the chiral fermionic zero mode in
the off-diagonal degrees of freedom. On the other hand, ref.~\cite{Pi}
calculated the effective potential between D0- and D8-branes, with the
approximation of small $b$ and $c_i$ ($b$ is the distance between D0- and
D8-branes along $X_9$, and $c_i$'s stand for backgrounds). Since the effective
potential is computed only for small $b$ and $c_i$, it is hard to determine
from the potential in~\cite{Pi} whether a fundamental string is created or not
when D0-brane crosses D8-brane.

The creation of a fundamental string in D0-D8 system can be related
to the creation of branes of other dimensions in other systems by sequences of
dualities. For example, after T-dualities, when one D4-brane along
(1357) directions crosses another D4-brane in (2468) directions, the
creation of a fundamental string occurs. But when two parallel D4-branes
cross each other, nothing happens. This raises an obvious question: is a
fundamental string created when two D4-branes at angles cross each other?

In the present paper, we compute the effective potential without the
approximation made in ref.~\cite{Pi} and use the result to discuss the brane
creation in M(atrix) theory. In particular we consider the creation of
a fundamental string (a longitudinal M2-brane in M-theory) when two D4-brane
(wrapped M5-brane in M-theory) at angles~\cite{BDL} cross each other.
The classical configuration we choose is the bound state
$\{(4+2+2+0)-(4+2+2+0)\}$, which is T-dual to two D4-branes at
angles~\cite{OZ}.

To describe the creation of a fundamental string, the one-loop effective
action for the present classical background is calculated, from which
the potential between two D4-branes at angles can be read off. It is
found that when the backgrounds $c_i$ are arbitrary, the effective
action for general $b$ is quite complicated, and the precise dependence of 
the potential on general $b$ cannot be determined. However, for
arbitrary $c_i$, the classical configuration $\{(4+2+2+0)-(4+2+2+0)\}$
breaks supersymmetry~\cite{BSS}, casting suspicion on the validity of
the one-loop approximation. When $c_1=c_2$, $c_3=c_4$ and $c_i\neq 0$,
on the other hand, the resulting configuration preserves $1/8$ unbroken
supersymmetry, and when $c_1=c_2=c_3=c_4\neq 0$, the unbroken supersymmetry
is enhanced to $3/16$~\cite{OZ,GGPT}. In these cases, we may rely
on our one-loop calculation. We find that when $c_1=c_2$, $c_3=c_4$ and
$c_i\neq 0$, the effective Lagrangian is surprisingly simplified, and is
given by $\frac{1}{2}\mbox{sign}(b)(b+a_0)$ (where we have turned on the
background $a_0$ for gauge field $A_0$), completely independent of
the backgrounds $c_i$. The resulting effective Lagrangian is anomalous under
the global gauge transformation~\cite{ERF,BSSi}, but is invariant under
$b\rightarrow-b$, $a_0\rightarrow-a_0$ transformation which corresponds
to the charge conjugation invariance~\cite{ERF}. One can
restore the invariance under the large gauge transformation by adding bare
Chern-Simons (CS) terms of the form $-\frac{1}{2}(b+a_0)$ to the effective
Lagrangian, but at the expense of breaking charge conjugation
invariance.\footnote{This could be understood as a result of
regularization~\cite{ERF}, or as bare terms allowed from symmetries in
the theory~\cite{BSSi}.}

We find the resulting potential in M(atrix) theory
is independent of the backgrounds $c_i$ related
to the angles between two D4-branes (after T-dualities). For
$c_1=c_2=c_3=c_4=0$, the potential vanishes, which indicates that when
two parallel D4-branes cross each other, no fundamental string is created.
For $c_1=c_2, c_3=c_4$, both nonzero, the potential produces a jump (exactly
the same amount of string tension) in the force acting on the brane when
the above two D4-branes at angles cross each other. In order to maintain the
BPS property, that is, for the total potential to vanish on both sides of
D4-brane, a fundamental string must be created. It is interesting to note
that this occurs independently of the nonzero angles between two D4-branes.
Our conclusion is consistent with the string calculations~\cite{DFK,BGL}.

Let us start with the M(atrix) theory Lagrangian which is the 10-dimensional
$U(N)$ super Yang-Mills Lagrangian reduced to $1+0$ dimensions~\cite{BFSS}
\EQ
\label{eq1}
L=\frac{1}{2g_s}{\rm Tr}\left(D_t X_i D_t X_i +\frac{1}{2}[X_i, X_j]^2
 +i\theta^\dagger D_t \theta +\theta^\dagger \gamma_i[X_i, \theta] \right),
\EN
where we have set $T_s^{-1}=2\pi\alpha'=1$, and $X_i$ $(i=1,\cdots,9)$ and
$\theta$ are bosonic and fermionic hermitian $N\times N$ matrices. 

We take the following background configuration:
\bea
\label{eq2}
&&\bar{A_0}=\left(
\begin{array}{ll}
 a_0 \mbox{\boldmath$1$} & 0 \\
 0     &0
\end{array}\right),\quad
\bar{X_1}=\left(
\begin{array}{ll}
 Q_1& 0\\
 0&0
\end{array}\right),\quad
\bar{X_2}=\left(
\begin{array}{ll}
 P_1& 0\\
 0& 0
\end{array}\right),\nn
&&\bar{X_3}=\left(
\begin{array}{ll}
 Q_2& 0\\
 0& 0
\end{array}\right),\quad
\bar{X_4}=\left(
\begin{array}{ll}
 P_2& 0\\
 0& 0
\end{array}\right),\quad
\bar{X_5}=\left(
\begin{array}{ll}
 0& 0\\
 0& Q_3
\end{array}\right),\quad
\bar{X_6}=\left(
\begin{array}{ll}
 0& 0\\
 0& -P_3
\end{array}\right),\nn
&&\bar{X_7}=\left(
\begin{array}{ll}
 0& 0\\
 0& Q_4
\end{array}\right),\quad
\bar{X_8}=\left(
\begin{array}{ll}
 0& 0\\
 0& -P_4
\end{array}\right),\quad
\bar{X_9}=\left(
\begin{array}{ll}
 b \mbox{\boldmath$1$}& 0\\
 0& 0
\end{array}\right),
\ena
where $[Q_1,P_1]=ic_1$, $[Q_2,P_2]=ic_2$, $[Q_3,P_3]=ic_3$, $[Q_4,P_4]=ic_4$.
Note that the background $a_0$ has been turned on for the gauge field $A_0$.
In the M(atrix) theory language, the D4-brane is described by a
configuration corresponding to a $U(N)$
instanton and this background configuration may be interpreted as
$(4+2+2+0)-(4+2+2+0)$~\cite{BSS}. After T-dualities $(T_{1357})$, it can also
be interpreted as one D4-brane lying in the $(1357)$ plane, the other being
rotated off the $(1357)$ plane by the rotations in the $(12)$, $(34)$, $(56)$
and $(78)$ planes~\cite{OZ}. The angles $\theta_{12},\theta_{34},\theta_{56}$
and $\theta_{78}$ mix the directions $(12)$, $(34)$, $(56)$ and $(78)$, and 
$\tan \theta_{12}, \tan \theta_{34}, \tan \theta_{56}$ and $\tan
\theta_{78}$ are proportional to $c_1, c_2, c_3$ and $c_4$, respectively.

In order to compute the effective Lagrangian, we expand Lagrangian (\ref{eq1})
to quadratic order in the fluctuations around the above background
($A_0=\bar{A_0}+Y_0, X_i=\bar{X_i}+Y_i$), and integrate out the off-diagonal
matrix elements which correspond to the degrees of freedom of the virtual
strings stretched between two D4-branes at angles.
The off-diagonal blocks can be chosen as~\cite{LM}
\EQ
\label{eq3}
Y_0=\left(
\begin{array}{ll}
 0& \phi_0 \\
 \phi_0^\dagger & 0
\end{array}\right),
Y_i=\left(
\begin{array}{ll}
 0& \phi_i\\
 \phi_i^\dagger & 0
\end{array}\right),
\theta=\left(
\begin{array}{ll}
 0& \psi\\
 \psi^\dagger & 0
\end{array}\right).
\EN
\vs{-3}
Following refs.~\cite{LM,CT,DKPS}, we can integrate out the off-diagonal modes
to get the following determinants (with $\tau=it$):
\bea
\label{eq4}
\mbox{Bosons:}
&&\mbox{det}^{-2}[-(\partial_\tau+a_0)^2+H]\nn
&&\mbox{det}^{-1}[-(\partial_\tau+a_0)^2+(H-2c_1)]
  \mbox{det}^{-1}[-(\partial_\tau+a_0)^2+(H+2c_1)]\nn
&&\mbox{det}^{-1}[-(\partial_\tau+a_0)^2+(H-2c_2)]
  \mbox{det}^{-1}[-(\partial_\tau+a_0)^2+(H+2c_2)]\nn
&&\mbox{det}^{-1}[-(\partial_\tau+a_0)^2+(H-2c_3)]
  \mbox{det}^{-1}[-(\partial_\tau+a_0)^2+(H+2c_3)]\nn
&&\mbox{det}^{-1}[-(\partial_\tau+a_0)^2+(H-2c_4)]
  \mbox{det}^{-1}[-(\partial_\tau+a_0)^2+(H+2c_4)],\\
&&\nn
\label{eq5}
\mbox{Ghost:}&&
\mbox{det}^2[-(\partial_{\tau}+a_0)^2+H],\\
&&\nn
\label{eq6}
\mbox{Fermions:}&&
\mbox{det}[(\partial_{\tau}+a_0)+m_f],
\ena
where
\bea
\label{eq7}
&&H=\sum_{i=1}^{4}(Q_i^2+P_i^2)+b^2,\nn
&&m_f=Q_1\gamma_1+P_1\gamma_2+Q_2\gamma_3+P_2\gamma_4-Q_3\gamma_5+
P_3\gamma_6-Q_4\gamma_7+P_4\gamma_8+b\gamma_9.
\ena

The $Q_i^2+P_i^2$ terms in $H$ are a collection of simple harmonic
oscillator Hamiltonians with eigenvalues $c_i(2n_i+1), n_i=0,1,\cdots$.
Since there is no tenth $16\times 16$ matrix which
anticommutes with all the $\gamma_i$'s, the fermionic determinant cannot
be converted into Klein-Gordon form by the usual method. We can still
compute $\frac{\partial\Gamma_F(b,a_0)}{\partial a_0}$ to find from (\ref{eq6})
\EQ
\label{eq8}
i \frac{\partial\Gamma_F(b,a_0)}{\partial a_0}
 ={\rm Tr}\frac{1}{\partial_{\tau}+a_0+m_f}.
\EN
Multiplying the denominator by $-(\partial_{\tau}+a_0)+m_f$ and using
Schwinger's proper-time representation, we have
\bea
\label{eq9}
i\frac{\partial\Gamma_F}{\partial a_0}&=&\mbox{Tr}\biggl( \left\{
-(\partial_{\tau}+a_0)+Q_1\gamma_1+P_1\gamma_2+Q_2\gamma_3+P_2\gamma_4
-Q_3\gamma_5+P_3\gamma_6-Q_4\gamma_7+P_4\gamma_8+b\gamma_9
\right\}\biggr.\nn
&&\biggl.\times
\int_{0}^{\infty}ds \exp\left\{-s\left[-(\partial_{\tau}+a_0)^2+H
+ic_1\gamma_{12}+ic_2\gamma_{34}+ic_3\gamma_{56}+ic_4\gamma_{78}
\right]\right\}\biggr).
\ena

Using the Fourier representation for the $\tau$ variable, one can take
its trace. After the shift of the integration variable, one finds that the
first term $(\partial_{\tau}+a_0)$ in the first curly bracket vanishes, and
the next 8 terms are zero because of the traces of odd numbers of $\gamma_i$.
We thus find that eq.~(\ref{eq9}) is reduced to 
\bea
\label{eq10}
i\frac{\partial\Gamma_F}{\partial a_0} =
\int_{0}^{\infty}ds \int_{-\infty}^{\infty}\frac{dk}{2\pi}
 \int d\tau b\mbox{Tr}\left(\gamma_9
\exp\left\{-s\left[k^2 + H + ic_1\gamma_{12}+ic_2\gamma_{34}
 +ic_3\gamma_{56}+ic_4\gamma_{78}\right]\right\} \right).
\ena
By choosing proper representation for $\gamma_i$, we can arrange
$\gamma_{12}, \gamma_{34}, \gamma_{56}, \gamma_{78}$ and $\gamma_{9}$ to 
take the form~\cite{GGPT}
\bea
\label{eq11}
&&\gamma_{12}=i\; \mbox{diag.}({\bf 1}_{8\times 8}, -{\bf 1}_{8\times 8}),\qquad
\gamma_{34}=i\; \mbox{diag.}({\bf 1}_{4\times 4}, -{\bf 1}_{4\times 4},
{\bf 1}_{4\times 4},-{\bf 1}_{4\times 4}),\nn
&&\gamma_{56}=i\; \mbox{diag.}(\overbrace{{\bf 1}_{2\times 2}, -{\bf 1}_{2\times
 2},\cdots, {\bf 1}_{2\times 2}, -{\bf 1}_{2\times 2}}^8), \quad
\gamma_{78}=i\; \mbox{diag.}(\overbrace{1, -1,\cdots, 1,
-1}^{16}),\nn
&&\gamma_9=\gamma_{12345678}=
\mbox{diag.}(1,-1,-1,1, -1,1,1,-1, -1,1,1,-1, 1,-1,-1,1),
\ena
where ${\bf 1}_{n\times n}$ is the $n\times n$ identity matrix.
Substituting (\ref{eq11}) into (\ref{eq10}) and performing the integrations
yield
\EQ
\label{eq12}
\frac{\partial\Gamma_F}{\partial a_0}=\int dt\frac{|b|}{2b},
\EN
for $c_i\neq 0$. When any one (two or three) of $c_i$ is (are)
zero, we find
\EQ
\label{eq13}
\frac{\partial\Gamma_F}{\partial a_0}=0.
\EN

It is easy to repeat a similar calculation for the contributions
from the bosons and ghosts, and we find that they vanish. Consequently
eqs.~(\ref{eq12}) and (\ref{eq13}) are the whole results for the derivative
of the total effective action:
\EQ
\frac{\partial\Gamma}{\partial a_0}=\int dt\frac{|b|}{2b},
 \quad \mbox{for $c_i\neq 0$},
\label{eq131}
\EN
and
\EQ
\frac{\partial\Gamma}{\partial a_0} = 0,
 \quad \mbox{when at least one of $c_i$ is zero},
\label{eq132}
\EN
with 
\EQ
\label{eq14}
\Gamma=\Gamma_B+\Gamma_G+\Gamma_F.
\EN
Here we point out that in deriving the results~(\ref{eq131}) and (\ref{eq132}),
we have not imposed any restriction on $c_i$.

Similarly, when $c_i\neq 0$, we find
\bea
\label{eq15}
\frac{\partial\Gamma}{\partial b}&=& \frac{b}{2\sqrt{\pi}}\int dt
 \int_{0}^{\infty}ds s^{-1/2}e^{-b^2 s}
 \sum_{n_1,\cdots,n_4=0}^{\infty} e^{-s\sum_{i=1}^{4} c_i(2n_i+1)} \nn
&&\hs{-3}\times\left\{
e^{-s(c_1+c_2-c_3-c_4)}+e^{-s(c_1+c_2-c_3+c_4)}+e^{-s(c_1+c_2+c_3-c_4)}
+e^{-s(c_1+c_2+c_3+c_4)}\right.\nn
&&\hs{-3}+e^{-s(c_1-c_2-c_3-c_4)}+e^{-s(c_1-c_2-c_3+c_4)}
+e^{-s(c_1-c_2+c_3-c_4)}+e^{-s(c_1-c_2+c_3+c_4)}\nn
&&\hs{-3}+e^{-s(-c_1+c_2-c_3-c_4)}+e^{-s(-c_1+c_2-c_3+c_4)}
+e^{-s(-c_1+c_2+c_3-c_4)}+e^{-s(-c_1+c_2+c_3+c_4)}\nn
&&\hs{-3}+e^{-s(-c_1-c_2-c_3-c_4)}+e^{-s(-c_1-c_2-c_3+c_4)}
+e^{-s(-c_1-c_2+c_3-c_4)}+e^{-s(-c_1-c_2+c_3+c_4)}\nn
&&\hs{-3}\left.-2e^{-2sc_1}-2e^{2sc_1}-2e^{-2sc_2}-2e^{2sc_2}
-2e^{-2sc_3}-2e^{2sc_3}-2e^{-2sc_4}-2e^{2sc_4}
\right\}.
\ena
Eq.~(\ref{eq15}) looks so complicated that it seems impossible to get
a definite result for $\partial\Gamma/\partial b$ without the approximation
for $c_i$ or $b$, as has been done in ref.~\cite{Pi}.\footnote{If one expands
eq.~(\ref{eq15}) in small $c_i$ up to fourth orders, one finds the result
reported in ref.~\cite{Pi}.}$^,$\footnote{
The first 16 terms in eq.~(\ref{eq15}) are the contributions from fermions.
We can easily see that there is only one zero eigenvalue in the spectrum
(in the thirteenth term) for $b=0$ and all $n_i=0$, which is the chiral
zero mode identified in ref.~\cite{HWu}.}
However, we are interested in the classical BPS-brane configurations in
M(atrix) theory with partially unbroken supersymmetry. For the present
classical configuration, we know that it has 1/8 unbroken supersymmetry
when $c_1=c_2$, $c_3=c_4$ and $c_i\neq 0$~\cite{OZ}. With this condition,
all the contributions from nonzero modes cancel out from eq.~(\ref{eq15}),
and we find $c_i$-independent result
\EQ
\label{eq16}
\frac{\partial\Gamma}{\partial b}=\int dt\frac{|b|}{2b}.
\EN
Note that when $c_1=c_2$, $c_3=c_4$, but $c_1$ or $c_3$ is zero,
eq.~(\ref{eq15}) reduces to 
\EQ
\label{eq17}
\frac{\partial\Gamma}{\partial b}=0.
\EN

{}From eqs.~(\ref{eq131})-(\ref{eq17}), the effective action for the
BPS-saturated background with $c_1=c_2$ and $c_3=c_4$ can be obtained:
\EQ
\label{eq18}
\Gamma(b, a_0)=0, \quad \mbox{for $c_1$ or $c_3$ is zero },
\EN
and
\EQ
\label{eq19}
\Gamma(b, a_0)=\int dt \frac{1}{2}\mbox{sign}(b)(b+a_0), \quad \mbox{for $c_1,
 c_3\neq 0$},
\EN
where the irrelevant integration constants (independent of $b$ and $a_0$)
have been dropped.

The potential obtained from the effective action~(\ref{eq19}) gives a jump
in the force at $b=0$ by a string tension. This has been interpreted in
terms of half strings in refs.~\cite{DFK,BGL}. We believe that the following
modification gives a better interpretation.

Note that eq.~(\ref{eq19}) possesses the charge conjugation invariance
under the transformation $b\rightarrow -b, a_0 \rightarrow -a_0$, but it 
is anomalous due to the presence of the term:
$\frac{1}{2}\mbox{sign}(b)a_0$. The factor $\frac{1}{2}$ indicates that
the partition function is not invariant under the global gauge
transformations, leading to a global anomaly~\cite{ERF,BSSi}.
The origin of this global anomaly is the chiral fermionic zero mode we
noted in the preceding footnote.
However, the invariance under the global gauge transformations can be
restored by adding bare CS terms to the Lagrangian, but they break
the charge conjugation invariance. From the viewpoint of the world-line
theory of the D-particle, when $b\ll l_s$, our D4-D4 system can be
approximately described by the open string model, and the resulting theory
is supersymmetric quantum mechanics with 4 supercharges. Imposed this
symmetry, the bare CS terms can be chosen as~\cite{ERF,BSSi,BGL}
\EQ
\label{eq20}
\Gamma_{cs}=-\int dt \frac{1}{2}(b+a_0).
\EN
Thus the effective Lagrangian is given by 
\EQ
\label{eq21}
{\cal L}_{eff}=\frac{1}{2}\mbox{sign}(b)(b+a_0)-\frac{1}{2}(b+a_0),
\EN
which restores the invariance under the large gauge transformation at
the expense of breaking charge conjugation invariance: $b\rightarrow -b$,
$a_0\rightarrow -a_0$. \footnote{A similar discussion of the effective action
is given in somewhat simplified manner for $N=8$ supersymmetric quantum
mechanics in ref.~\cite{BSSi} and is later exploited in refs.~\cite{BGL,HWu}.}
Actually this modification can be interpreted as different choices of the
regularization schemes~\cite{ERF}, for example, using Pauli-Villars
regularization.

We also note that the theory is anomaly-free on the backgrounds of D0-brane
interacting with D2-, D4- and D6-branes, as can be seen from eq.~(\ref{eq132}).

Performing T-dualities ($T_{1357}$), the $\{(4+2+2+0)-(4+2+2+0)\}$
background configuration can be interpreted as one of the D4-brane lying on
$(1357)$ plane, and the other rotated away from $(1357)$ plane along
$(12)$ and $(34)$ directions with an angle $\theta_1$ and along $(56)$
and $(78)$ directions with an angle $\theta_2$~\cite{OZ}.
Under T-dualities mapping the configuration to that of two D4-branes at
angles, the distance $b$ is not changed. Hence after choosing the $a_0=0$
gauge, the interaction potential between two D4-branes at angles at distance
$b$ is simply given by~\cite{Li}
\EQ
\label{eq22}
\Gamma(b, a_0=0)=-\int dt V(b).
\EN
{}From eqs.~(\ref{eq21}) and (\ref{eq22}), the effective potential between
two D4-branes at angles can be read off as
\EQ
\label{eq23}
V(b)=-\frac{1}{2}T_s(|b|-b),
\EN
where we have switched on the string tension $T_s=(2\pi\alpha')^{-1}$.
An important point is that the jump in the force at $b=0$ is
unchanged by this procedure.

We emphasize that the jump in the force is due to the contribution of the
fermionic zero mode. Also note that the potential in eq.~(\ref{eq23}) is
independent of the angles between two D4-branes, and is the same as that
for two orthogonal D4-branes with eight ND-directions.

Eq.~(\ref{eq18}) shows that when $c_1$ or $c_3$ is zero, the potential
vanishes. Suppose $c_3=0$. We have the following physical picture: the first
D4-brane lies on $(1357)$ directions, and the second is rotated off (13)
plane along (12) and (34) directions with two common directions (57)
with the first D4-brane. This configuration corresponds
to two D2-branes at angle and the rotation is a real element of $SO(2)$
\cite{BDL}. Since the potential vanishes, when two such D4-branes cross
each other, no fundamental string is created.

Let us next consider $c_1=c_2=c_3=c_4=c$ case where the unbroken
supersymmetry is enhanced form $1/8$ to $3/16$~\cite{OZ,GGPT}, and in the
resulting configuration the orientations of the two
D4-branes are related by a rotation in $Sp(2)$ subgroup of $SO(8)$
commuting with multiplication by a quaternion~\cite{GGPT}. If $c=0$ 
i.e., two D4-branes are parallel and lie on $(1357)$ directions, the
unbroken supersymmetry is further enhanced from $3/16$ to $1/2$
\cite{Po,PP}. Since the potential vanishes, no fundamental string is
created when two parallel D4-branes cross each other. If $c\neq 0$, the
potential gets contributions only from the fermionic zero mode (and bare CS
terms) and is given by eq.~(\ref{eq23}) which shows that there is no force on
one side and a repulsion of a single string tension occurs when two D4-branes
at angles cross. This
force is canceled by a fundamental string created between them in order
to maintain the BPS property. The origin of this mechanism will be further
discussed below. Since eq.~(\ref{eq23}) is independent of $c_1(c_3)$,
the above conclusion is valid for finite angles. In particular the case
of the angles equal to $\frac{\pi}{2}$ corresponds to two orthogonal
D4-branes with 8 ND-directions related to that extensively discussed in the
literature~\cite{HW,BDG,DFK,BGL,dA,HWu}.

We thus find that the potential obtained from the effective Lagrangian in
M(atrix) theory misses the term from $R(-1)^F$ sectors,\footnote{
A similar observation was also made in refs.~\cite{BGL,Pi}.} but
it can be modified to agree with that obtained
from the string calculations in~\cite{BGL}.

{}From the above discussion, we see that the brane creation is closely
related to the existence of a chiral fermionic zero mode in the off-diagonal
degrees of freedom, which in the present case manifests itself in the
form of the global anomaly, and the created string is needed to cancel
this global anomaly completely. This can be understood from the point of
view of the worldvolume theory. Consider two D4-branes:
D${}^{(1)}$4 and D${}^{(2)}$4 with their worldvolumes $B_1, B_2$, and
the intersection between them $B_{12}=B_1\cap B_2$ is I-brane~\cite{GHM}.
The worldvolume actions of D${}^{(1)}$4, D${}^{(2)}$4 and I-brane are
denoted by $S_1, S_2$ and $S_I$. Under a gauge transformation, the
variation of $S_1$ and $S_2$ has a boundary piece localized at $B_{12}$,
which precisely cancels the anomalous variation of $S_I$~\cite{GHM,BDG}.
In our case, the effective action calculated from the off-diagonal
blocks in M(atrix) theory corresponds to the action $S_I$ of the
worldvolume theory. Under large gauge transformation, the effective
action has a global anomaly indeed. Such anomaly from $S_I$ can be
canceled by the variation of the bulk actions. Thus the inflow of
charge that is required for the absence of global anomaly can be regarded
as the created fundamental string.

In M(atrix) theory, to implement such anomaly cancellation, one would add
explicit CS terms $\int C\wedge \mbox{tr} e^F$ to the M(atrix) theory
Lagrangian, and the anomaly can be canceled by the anomaly inflow from
the bulk mediated by these CS terms~\cite{Ho,KR}. Indeed, such CS terms
combined with the D8-brane background were used in ref.~\cite{DFK} to
cancel the potential. In~\cite{Ho}, such CS couplings were suggested
to be an effective description of a more microscopic mechanism where
the supergravity background could be generated by integrating out certain
heavy matrix modes. The effective result could be expected as the source
for the creation of the fundamental string and extra degrees of freedom
corresponding to the created string might be found from
D${}^{(1)}$4-D${}^{(1)}$4 and D${}^{(2)}$4-D${}^{(2)}$4 strings. On the other
hand, the recent work in refs.~\cite{SDR} suggests that there may be more
degrees of freedom to M(atrix) theory than just 0-branes. It would be
interesting to see if these CS terms have any implications for ``corrections"
to the original model~\cite{BFSS}. Work along this line is under investigation.

In conclusion, the creation of a fundamental string when two D4-branes
at angles cross each other has been discussed in the context of M(atrix)
theory. When $c_1=c_2$, $c_3=c_4$ and $c_i\neq 0$, the background
possesses $1/8$ unbroken supersymmetry, the effective action is
surprisingly simplified.
We have found that the potential obtained from the effective Lagrangian
possessing the invariance under the large gauge transformation is independent
of $c_i$ related to the angles between two D4-branes (after T-dualities)
and exhibits a jump in the force by the amount of a string tension.
This result indicates that a fundamental sting is created indeed when two
D4-branes at angles (preserving $1/8$ or $3/16$ unbroken supersymmetry) cross
each other in order to cancel the force. Such a string creation is related
to the existence of the global anomaly, and we have interpreted the created
string as the object required to cancel this global anomaly completely.
Our results are consistent with string calculations~\cite{DFK,BGL}.

\vs{5}
\noindent
{\bf Acknowledgments:}
We would like to thank I. Klebanov and T. Nakatsu for helpful comments.
This work was supported in part by Grand-in-aid from the Ministry of
Education, Science, Sports and Culture No. 96208. J.-G. Zhou thanks the
Japan Society for the Promotion of Science for the financial support.

\newcommand{\NP}[1]{Nucl.\ Phys.\ {\bf #1}}
\newcommand{\AP}[1]{Ann.\ Phys.\ {\bf #1}}
\newcommand{\PL}[1]{Phys.\ Lett.\ {\bf #1}}
\newcommand{\NC}[1]{Nuovo Cimento {\bf #1}}
\newcommand{\CMP}[1]{Comm.\ Math.\ Phys.\ {\bf #1}}
\newcommand{\PR}[1]{Phys.\ Rev.\ {\bf #1}}
\newcommand{\PRL}[1]{Phys.\ Rev.\ Lett.\ {\bf #1}}
\newcommand{\PRE}[1]{Phys.\ Rep.\ {\bf #1}}
\newcommand{\PTP}[1]{Prog.\ Theor.\ Phys.\ {\bf #1}}
\newcommand{\PTPS}[1]{Prog.\ Theor.\ Phys.\ Suppl.\ {\bf #1}}
\newcommand{\MPL}[1]{Mod.\ Phys.\ Lett.\ {\bf #1}}
\newcommand{\IJMP}[1]{Int.\ Jour.\ Mod.\ Phys.\ {\bf #1}}
\newcommand{\JP}[1]{Jour.\ Phys.\ {\bf #1}}

\end{document}